*Sergey Andreyev*

# User-driven applications – new design paradigm

Programs for complicated engineering and scientific tasks always have to deal with a problem of showing numerous graphical results.  The limits of the screen space and often opposite requirements from different users are the cause of the infinite discussions between designers and users, but the source of this ongoing conflict is not in the level of interface design, but in the basic principle of current graphical output: user may change some views and details, but in general the output view is absolutely defined and fixed by the developer.  Author was working for several years on the algorithm that will allow eliminating this problem thus allowing stepping from designer-driven applications to user-driven.  Such type of applications in which user is deciding what, when and how to show on the screen – this is the dream of scientists and engineers working on the analysis of the most complicated tasks.  The new paradigm is based on movable and resizable graphics, and such type of graphics can be widely used not only for scientific and engineering applications.

**Introduction**

For the last 20 years users of personal computers are working on two levels.  On the upper level the space of one or several screens is shared by all the applications; any started application will show a piece of information inside the rectangular window.  Each window is movable and resizable; we can move them around the screen, make them bigger or smaller, put them side by side or overlap.  At any time each user decides about the placement and the size of any window.  This is **axiom 1** in modern day programming design: <u>on the upper level all windows are movable and resizable</u>.  To make these features obvious the upper level windows have title bars (to grab and move the whole window) and borders (to move any one of them thus resizing the window). These features – movable and resizable – are the standard features of any window, and only for special purposes those features can be eliminated.

By starting any application we enter another (inner) level with slightly different, but also strict rules: on that level we have two different types of objects - controls, that inherit everything from windows, and graphical objects that have no inheritance from them and are absolutely different.  The inheritance of controls from windows is not obvious to the users, as controls often do not look like windows; usually there are between 20 and 30 different basic controls plus variations.  Controls have no title bars, so there is no indication, that they can be moved; usually there are no such borders that inform about the possibility of resizing.  But these inherited features of all controls (movable and resizable) can be easily used by the programmers, for example, in the form of anchoring and docking.  To use these features or not is defined by the designer of the program, and if it is allowed, then more than often it is done indirectly.  It is organized through one or another form of interface design, every two or three years the most popular word for such form of design will change; just now the most popular term is adaptable interface.  But the most important thing is not *how* it can be done, but that for the controls moving and resizing *can be done* without problems.

Graphical objects are of absolutely different origin than controls and by default they are neither movable, nor resizable.  There are always some tricks to make things look not what they are in reality (the programmers are even paid for some knowledge of such tricks), and one of the often used tricks is painting on top of the control: any panel is a control, so it is resizable, and with the help of anchoring / docking features it is fairly easy to make an impression as if you have a resizable graphics.  By default panels have no border, and if the back color of the panel is the same as of the parent window, then there is no way to distinguish between painting in the window, or on the panel which resides on it.  Certainly, such "resizable" graphics is only an imitation, but in some cases it is just enough.  All depends on the purpose of application.  Another solution for resizing of rectangular graphical objects is the use of bitmap operations, but in most cases it can't be used because of quality problems, especially, while enlarging the image. Both of these tricky solutions (painting on panel or using bitmap operations) have one common defect – they can be used only with rectangular objects.

If any limited area is populated with two different types of tenants (controls and graphical objects) that prefer to live under different rules, then the only way to organize their peaceful residence and avoid any mess is to force them to live under ONE law. Because the currently used graphics is neither movable, nor resizable, then the easiest solution is to strip all controls of those features. That is also the reason, why the percent of applications that are allowing users to move around any inner parts is so small.

Thus we have **axiom 2**: <u>on the inner level the objects are usually neither movable, nor resizable</u>.

What is interesting, that these two axioms bring us to the absolutely paradoxical situation:

- on the upper level, which is not so important for real work, any user has an absolute control of all the components, and any changes are done easily;
- on the inner level, which is much more important for any user, because the real tasks are solved here, user has nearly no control at all, and if he has, then it is very limited and is always organized indirectly through some additional windows.

What is amazing, that this paradoxical situation is really looked at as an axiom, so I can't remember (and can't find) even a single paper that will doubt the fairness of this situation and will show an attempt to change axiom 2 to something better, but there are tons of papers and numerous discussions about the better way of organizing new interface for dealing with the parameters of graphical objects. I have no doubt in the importance of such discussions, as I am working on interface design for many years, but there are very vast areas of applications where the question of finding an easy way of moving and resizing of graphical objects is much more important than anything else, and if such an easy algorithm would be found, then it may have very interesting and far going consequences. Even the whole programming of such applications can be changed to the huge benefit of the end users.

Axioms I mentioned were never declared as axioms in strict mathematical way; at the same time I never saw, read or heard even about a single attempt to look at this not as an axiom and to design any kind of application on different foundation.

Millions of users are living with these two axioms for years, and there are no questions of fairness of it. To understand the paradox of this situation look at the similar one projected on our everyday life: we have an ability to move freely around the city, but we would have absolutely strict rules on how and where each piece of furniture can be and must be placed in our houses. And no questions about such ruling?

Author is designing the systems for extremely complicated areas (speech recognition, telecommunication, thermodynamics, electricity network analysis) for many years, each one of those systems had a lot of graphical output, and so each one had to deal with the problem of limited output area and unlimited demand on the number of plots. Just recently I found the solution for movable and resizable graphics and though for years I was looking for some solution mainly for rectangular areas (plots are always rectangular), the result is used now for graphical objects of absolutely different shapes. Before I'll describe how it can be done and, what is much more important, the consequences of using this new graphics I want to emphasize that:

1. Though the main goal was to find the solution for plotting, the algorithm (and classes) can be used far away from the described area.
2. The algorithm (and classes) must be (and can be) looked on separately from the consequences of using them. Classes can be organized in different way, algorithm may be also different, but if there is any form of movable and resizable graphics, then it is the base for absolutely new paradigm – user-driven applications.

3. It is not simply an idea of "it would be nice to have" type. The classes and algorithm of new plotting were designed and are working now; the results can be seen and checked at http://www.doublesharp.org/TuneableGraphics.htm

4. Everyone is welcomed not only to look at but to try the new algorithm for any kind of projects. In http://www.doublesharp.org/Test_MoveGraphLibrary.zip there is the full testing solution with all codes and DLL. In http://www.doublesharp.org/Test_MovableResizableGraphics.doc there is the short description of that project and few words about initializing and using this movable graphics. In http://www.doublesharp.org/MoveGraphLibrary_Classes.doc there is the description of classes, so you can do much more than is shown in this test program (I tried to make it really small and very obvious).

**User-driven applications – the new design paradigm**

Let's assume that there is such graphics that allows any user to move and resize any plot on the screen at any moment while the application is running. How (if at all) this can affect the design process and how it can affect the user's work – the main purpose of any engineering (scientific) application? To analyze the situation let's look at the well known case of an application for analysis of mathematical functions.

There are several well known programs or packages which allow viewing of variety of mathematical functions or will show the plot(s) based on previously calculated data. Each of the programs has its own interface for setting the viewing parameters and determining the number of shown plots and their placements. Each program sets some limit on the number of shown plots (one or several), on their placement and size (in the same single area; side by side either vertically or horizontally; in several rows and columns). It doesn't matter how many plots are allowed to be shown simultaneously and how many choices the user has to place the plots on the screen – each case is predefined by the developer of the program and for each case you can say without running the application where and how each plot would be shown. This is an example of *designer - driven application*, and not only this classical one but all applications for engineering and scientific tasks are developed in this and only this way. Users are always working inside the scenario written by the designer to the tiny details.

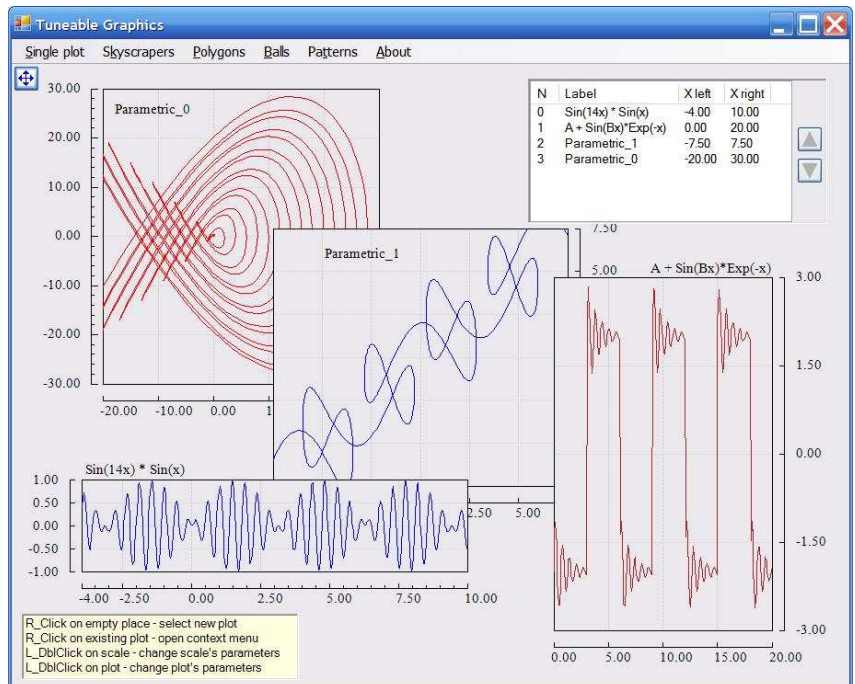

Figure 1. Main view with combination of plots

For user-driven applications it would be absolutely different. On figure 1 is shown the standard view of previously mentioned application, when it is started. It looks like an ordinary view of several partly overlapping plots, only user can grab, move and resize each of them at any moment; the new plots can be added and the existing can be deleted. The group of controls (list and two buttons on the panel) can be also moved around the screen.  The special lines and areas (connections and nodes – I'll explain them later) which are sensitive and allow moving and reconfiguring can be switched ON / OFF by the small button in the upper – left corner, but while moving the mouse around the screen the changing of the mouse cursor will signal that reconfiguring can be done even without visualizing those special lines.

What makes this application absolutely different from standard designer-driven applications that it is impossible to predict the view of the application beforehand.  User and only user has an absolute control of **what, when** and **how** must be shown at any moment; this is the main idea of the *user-driven applications*.  And it also means that in this type of applications (on the inner level) user gets the same degree of flexibility as up till now he has only on upper level.  This is absolutely new in design of applications.

The developer of such program is not designing the output view, but has to give to the user an instrument to put on the screen whatever he –USER - wants.  That is the dream of any user of any engineering or scientific program: to have at the screen at any moment only what he really needs and in the way he prefers to see it.  So what was changed in the main design and what makes it possible to move from designer-driven applications to user-driven?

Such transformation is possible if and only if we have movable and resizable graphics.

**Movable and resizable graphics**

My idea of making graphical objects movable and resizable is based on **contour presentation**.  (Contour looks like something very familiar from the graph theory, but the standard words from this theory often conflict with the meaning of the same words used by those who are programming under Windows for many years, so I'll try to use different words with obvious meanings.)  Any graphical object that needs to be resizable and / or movable will have a contour consisting of nodes and connections.  Nodes are used as sensitive areas which can be moved separately, and by moving any one of them the sizes and the shape of the whole object can be changed.  Connections are used for grabbing and moving of the whole object.  Sizes and forms of the sensitive nodes can vary and the same with the sizes of sensitive areas (width) around connections.

The important feature: contours do not necessarily copy the shape of the object.  Contours and nodes have to organize an easy way of resizing and moving the object and for this purpose they are taking under their control some part of the screen; at the same time they have to be organized in such a way as not to interfere with all other needed mouse clicks aimed at changing other parameters of the graphical objects.  For example, in all standard XY plots (figure 1) double click on the main area of any of them will open special window for setting of the area's parameters, and because of this the contour is moved slightly out of the main plotting area.  For XYZ plotting the contour's sensitive area consists of the narrow strips around axes.

When the designer decides that moving / resizing would be useful parameters of some type of graphical objects, then he organizes the contour taking into consideration the shape of the object, possible transformations and the set of already used mouse clicks (if there are any) associated with such objects.  Two other main functions for moving the whole object and reconfiguring (moving some of the nodes separately) depend on how this moving / resizing is planned to be used.

If the graphical object has to be unmovable but resizable, then there will be no connections and only nodes.  If there must be some restriction even on positioning of those nodes, it is done easily with the clipping.  If on the contrary moving is allowed but not resizing (no general shape changing), then the contour will have empty nodes, but the set of connections.  This very flexible combination of nodes and connections covers all the possible scenarios, but in each particular case the contour is based on the type and form of the graphical object and how you need to work with it.  For some objects the shape of the contour can be the same as an outer frame of an object, but this is a rare situation; in general contour can be of absolutely different shape and looks more like a skeleton.  I especially included into the application different types of contours to show the wide variety of objects and different types of applications where this algorithm can be used.

Identical rules are applied not only to graphical objects, but to controls or groups of controls as well. This is one of the main ideas: similarity in behavior of all the applications' parts without any division on graphical objects and controls makes it easier for any user to deal with the application. There can be some limitations set by the developer, but these limitations would be based only on the ideas of best design: I don't think that users would be happy to move around the screen each control separately, especially if there are a lot of them, but if there would be an easy way to move around the whole group of controls and organize the view in the best way for each particular user – this would be an extremely useful feature of any application. For example, in the application shown on figure 1, there are no problems at all to make movable separately the list with the information and two buttons at the side of it, but because these buttons are used for changing the order of items in the list (and thus changing the order of plots drawing) then the best place for the buttons is always at the side of the list, so it makes more sense to combine them and to give them one contour, thus allowing their simultaneous movement.

The whole algorithm is very easy in use. First, you initialize the Mover, which will store all movable objects
```
Mover Movers = new Mover ();
```
Then add to it every object (and not only graphical), that you would like to be movable. For example, in the case of the above window while starting the application I added the panel, on which the list and two buttons reside.
```
MovableObject mover = new MovableObject (panelWithList);
Movers .Add (mover);
```
When I want to put on the screen the plot of the new function, I am doing exactly the same
```
DemoFunction demoFunc = new DemoFunction (…);
Movers .Add (new MovableObject (demoFunc));
```
Now the Movers has everything to do the whole job, and I need only to include three handles.

To grab any object by L_Press
```
private void OnMouseDown (object sender, MouseEventArgs mea)
{
    if (mea .Button == MouseButtons .Left)
    {
        Movers .CatchMover (mea .Location);
        …
    }
}
```

To move already grabbed object or inform that something underneath the mouse cursor can be grabbed
```
private void OnMouseMove (object sender, MouseEventArgs mea)
{
    Movers .MovingMover (mea .Location);
}
```

To release the object, when the mouse button is released
```
private void OnMouseUp (object sender, MouseEventArgs mea)
{
    Movers .ReleaseMover ();
}
```

Because there are no limitations on the forms and sizes of nodes and the complexity (or simplicity) of connections this algorithm can be used for making movable / resizable a wide variety of objects. Let's look at some samples which are far away from the standard rectangular shape.

**Samples of movable and resizable objects**

The first nonrectangular object on which I was checking the new algorithm was XYZ coordinate system. To make it more interesting I added the type of plot, which is never used in engineering projects, but very often in financial analysis. I call it *skyscrapers*, though there can be some special word (Fig.2). Depending on the set of data there is a very high probability that at any fixed angles of axes some of the *towers* (or a lot of them) can be closed from view by the higher towers in front. With the axes' system movable and resizable (you can see those four sensitive nodes and their connections) for any combination of data the user will immediately set the best angle and size for viewing as much data as possible. Or he can turn the whole picture around to look at the parts that are closed from observation on different angle.

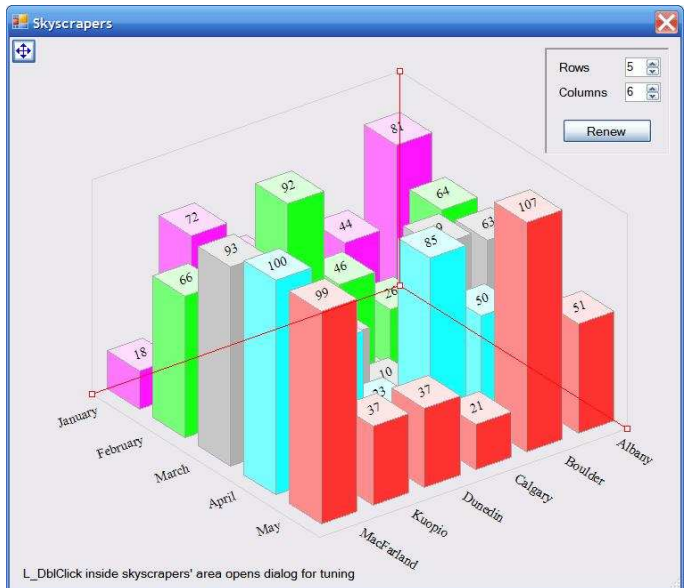

Figure 2. Skyscrapers with nodes and connections

This is also a good example of interconnection between the graphical object and its contour. For XYZ plotting it would be the obvious decision to organize a contour as a copy of the axes. At the same time the user of such skyscrapers' plot can go into special window for tuning the skyscrapers by double clicking anywhere inside their area. By adding this movable / resizable feature the designer excludes the thin strips around axes from the area which will open the tuning window, so this is the designer's responsibility to define the width of these strips (though it can be easily given to the user as another parameter).

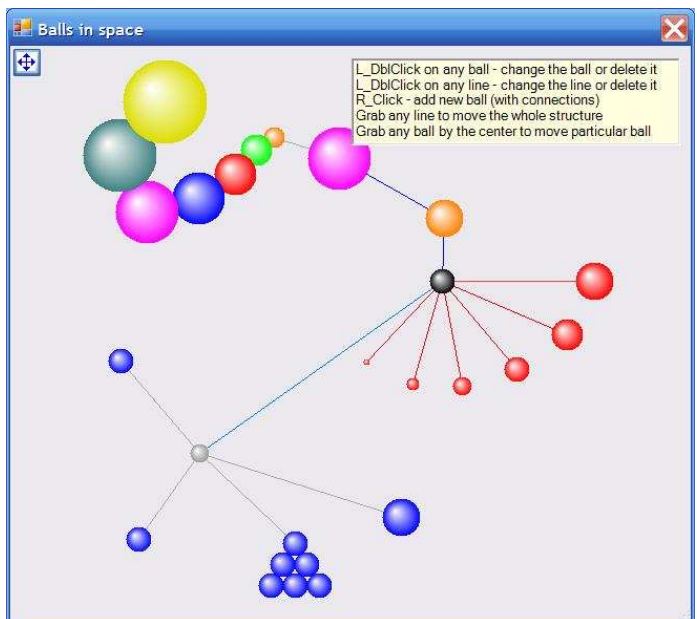

Figure 3. Balls in space

Another sample is of absolutely different shape and area; similar pictures are often seen in applications for chemistry or medical projects (Fig.3). No limitations on number of balls or connections; users receive an instrument to design any structure consisting of *balls* and connections. They can be added, deleted, changed. In this case it looks like an obvious thing to define the contour as the copy of the graph, but this is not compulsory.

The situation with movable but not resizable objects is very widely used. In the mentioned application there is an example of one logical game, which is using this algorithm to move the tiles around the screen. The same algorithm can be used in a wide variety of design programs dealing with the best placement of graphical objects of arbitrary form inside some fixed area.

These samples also illustrate another aspect of the used algorithm. Sample on figure 4 includes the big number of independent graphical objects; the sample from figure 3 includes only one object, but it can consist of huge number of elements; on figure 1 the number of objects is not limited and the complexity of each object can vary. It doesn't matter how many objects are there and what is the complexity of each of them; all of them are treated in the same way with the handles that were mentioned above.

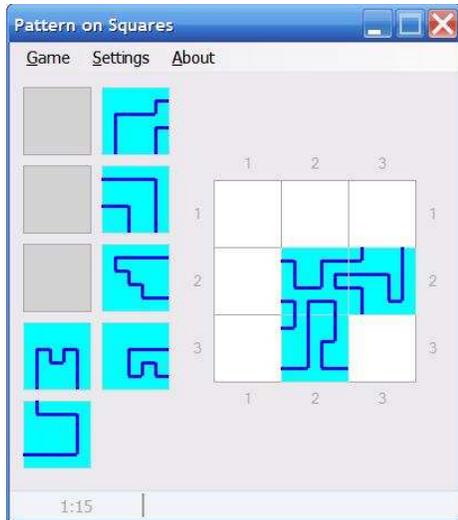

Figure 4. Logical game

Let's summarize some features of this extremely flexible contour design (nodes plus connections), which from my point of view can cover any possible situation to organize moving and resizing of the objects.

- Contour can consist of any number of nodes and their connections.
- Nodes can be moved individually thus allowing to reconfigure the object; by grabbing any connection the whole object can be moved (fig. 1).
- Contour may consist of a serious of connections between empty nodes. That makes an object movable, but not resizable (fig. 1; several controls on top of the panel).
- Each of the nodes has its own parameters, and by connecting nodes of different types it is easy to allow, for example, resizing along one direction but prohibit along another. This means organizing limited reconfiguring. (Such type of objects – scale - is included into application, but is not shown here.)
- It doesn't matter that some contours may represent graphical objects and others controls or groups of controls. All contours are treated in the same way, thus allowing the user to change easily the inner view of any application.

**New graphics, new design, new possibilities**

There are two different parts in these described new results: graphics itself and applications, based on it. I am emphasizing the difference between these things as not always they are strictly linked with one another.

The kernel of the new things is definitely the new graphics. I constructed it on the base of contour presentation so, that it can cover any possible demand for moving and resizing objects on the screen. At the same time this new technique doesn't substitute anything that was used before; it is simply added as a very powerful new feature, so I simply added it to the graphical classes that I was using for at least a decade.

As any other feature it can be used or not used, and the switch ON / OFF of this feature can be done in the same easy way as with all other features. (Don't confuse it with switching ON / OFF the nodes and connections in the mentioned application; these are absolutely different things.) Moving and resizing is simply the feature of any object, so there is no demand that either all objects must be movable or none.

The moving / resizing feature can be used by itself without any changes in design of the whole application. It would be like improving the interface, but the application will be still designer-driven. There are a lot of applications which do not need any more changes beyond this.

But there are a lot of applications, especially very complicated applications with huge amount of output plotting, which can be significantly changed with the use of such graphics. From author's experience this would be the programs for engineering tasks and scientific research; there can be other vast areas, but now let's look at this type of applications. Just now there are mostly two types of output: either the graphical output is organized by designer or users take numeric results and put them into additional program to turn them into plot. The use of new graphics can be a real revolution in design of such complicated engineering (scientific) applications, but to receive the best results designers would have to switch from controlling the output view of an application to giving to the users the whole access to all results and producing them an easy mechanism of visualizing those results in any way and in any place. Thus shifting from designer-driven to user-driven applications. The movable / resizable graphics can be achieved in different ways, but user-driven applications can be based only on such type of graphics.

The consequences of using the movable graphics are definitely much more important than simply adding one new feature to existing graphics. And it is nearly impossible to estimate all these consequences without starting to use such graphics and writing the programs which will especially benefit on such feature. To estimate the difference between simply new feature and consequences of using it lets look at another one (from my point of view very similar) situation.

We are living in the houses that are fixed on some pieces of property and have fixed outside sizes and inner planning. Some changes can be made from time to time, but they need a lot of efforts and money. And in cases when we really need bigger or smaller house we are usually forced to relocate, which is not always the best thing to do but always have a lot of sociological consequences (discussion of them is not the purpose of this paper.) Now there is the new unknown company in the market, which is selling the new adjustable buildings, advertising that WITHOUT ANY EXTRA COST you can receive (at the same price!) the house which you can stretch (or shrink, if you want some extra place around) at any moment. Just touch the wall with the wand and tell it to move. (Again let's not mention here what the solid well known constructors would like to do with the new developer. This is not the theme for discussion in programming.) I am absolutely sure that at the beginning 110 percent of the people would be suspicious. The ordinary questions would be "And what is going to happen to all the pipes and wires?", "If I want to move the wall, what am I going to do with my flowerbed?" The explanation of that construction company, that everything is going to be fine, exactly what they want, will not lessen the suspicions. But the point that I want to mention here is not the benefits of the family who would be courageous enough to sign the contract with the new developer. The benefits would be certainly huge, but the consequences would be much wider. Just try to think about it for a minute. (This is certainly out of this paper's discussion.) If we would have such construction, it would change our whole life, the life of the society as a whole. The consequences are definitely much more important than the mechanism that allowed to achieve them.

I think that movable / resizable graphics can change a lot in the design of programs.

*Conclusion*. In the modern world of programming we have user-driven very flexible system of windows on the upper level and absolutely designer-driven inner level, on which users are doing the real work. The flexibility of this level is not only very low; for the most complicated engineering and scientific applications this level is hardly adjustable to the users' demands.

The cause of this awkward situation is the use of the fixed graphics, which is absolutely defined by the designer at the stage of development. Author designed an algorithm, based on contour presentation of graphical objects, which makes arbitrary graphical objects movable and resizable. The same algorithm is used both with graphical objects and controls; as the result applications, based on such algorithm, receive the same level of flexibility as we have now only on the upper level. This means not only working with better graphics, but much more important that the new graphics is the base of the new paradigm – user-driven applications.

In such type of applications user has an absolute control of **what, when** and **how** must be shown at any moment. Changing to such type of graphics is easy, but to get all the benefits will require some changes in the design ideas. But the improvements in the output of applications from the users' point of view can be enormous and their ability to analyze the most complicated and interesting cases will switch to another, much higher level.

**Acknowledgments**

Many thanks to those with whom I was discussing for years the design of tuneable graphics, beginning with Dr. Stanislav Shabunya from the Heat and Mass Transfer institute. And also to those with whom I was discussing the new movable graphics through the last year. Special thanks to those who didn't get the new ideas from the beginning and forced me into better and better explanations and further work on the algorithm.

Dr. Sergey Andreyev  ( [andreyev_sergey@yahoo.com](mailto:andreyev_sergey@yahoo.com) )